 \documentclass[lettersize,journal]{IEEEtran}
\usepackage{amsmath,amsfonts}
\usepackage{algorithmic}
\usepackage{algorithm}
\usepackage{graphicx}
\usepackage{array}
\usepackage{textcomp}
\usepackage{stfloats}
\usepackage{url}
\usepackage{verbatim}
\usepackage{multirow}
\usepackage{subfigure}
\usepackage[table]{xcolor}
\usepackage{cite}
\usepackage{array}
\usepackage{makecell}

\usepackage{amssymb}
\usepackage{amsmath}
\begin{document}


\title{Derivative-Free Optimization-Empowered Wireless Channel Reconfiguration for 6G}
\author{Peilan Wang, Jun Fang, Xianlong Zeng, Bin Wang, Zhi Chen,
and Yonina C. Eldar, ~\IEEEmembership{Fellow,~IEEE}
\thanks{Peilan Wang, Jun Fang, Xianlong Zeng, and Zhi Chen are with the National Key Laboratory of Wireless Communications, University of
Electronic Science and Technology of China, Chengdu, 611731, China, Email: JunFang@uestc.edu.cn}
\thanks{Bin Wang is with the Test and Training Base, National University of Defense Technology, Xian, China, 710106}
\thanks{Yonina C. Eldar is with the Faculty of Mathematics and Computer Science, Weizmann Institute of Science, Rehovot, 7610001, Israel. Email: yonina.eldar@weizmann.ac.il
}}

\maketitle



%

\begin{abstract}
Reconfigurable antennas, including reconfigurable intelligent surface (RIS), movable antenna (MA), fluid antenna (FA), and other advanced antenna techniques, have been studied extensively in the context of reshaping wireless propagation environments for 6G and beyond wireless communications. Nevertheless, how to reconfigure/optimize the real-time controllable coefficients to achieve a favorable end-to-end wireless channel remains a substantial challenge, as it usually requires accurate modeling of the complex interaction between the reconfigurable devices and the electromagnetic waves, as well as knowledge of implicit channel propagation parameters. In this paper, we introduce a derivative-free optimization (a.k.a., zeroth-order (ZO) optimization) technique to directly optimize reconfigurable coefficients to shape the wireless end-to-end channel, without the need of channel modeling and estimation of the implicit environmental propagation parameters. We present the fundamental principles of ZO optimization and discuss its potential advantages in wireless channel reconfiguration. Two case studies for RIS and movable antenna enabled single-input single-output (SISO) systems are provided to show the superiority of ZO-based methods as compared to state-of-the-art techniques. Finally, we outline promising future research directions and offer concluding insights on derivative-free optimization for reconfigurable antenna technologies.

\end{abstract}


\begin{figure*}[!t]
\centering
\includegraphics[width=7.in]{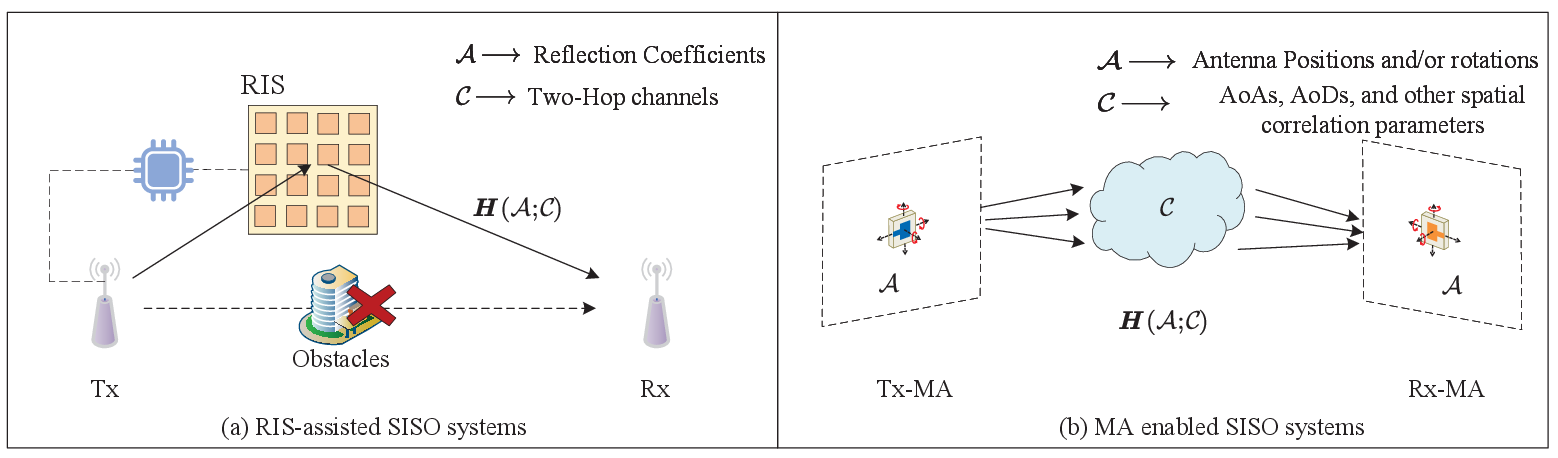}
\caption{Illustration of RIS-assisted systems and MA enabled systems. In (a), the adjustable parameters $\mathcal{A}$ are the reflection coefficients of the RIS, while the propagation parameters $\mathcal{C}$ represent the Tx-RIS and RIS-Rx channels. In (b), the parameters $\mathcal{A}$ are antenna positions and orientations, while the typical propagation parameters for MA communication systems include attenuation, angular and delay parameters associated with each signal path.}
\label{Fig-H1}
\end{figure*}



\section{Introduction}
Facing 2030 and beyond, sixth-generation (6G) wireless communications aim to surpass 5G by enhancing key performance indicators by at least two orders of magnitude, while introducing innovations for ubiquitous connectivity \cite{liuSeventyYearsRadar2023}. To achieve these ambitious goals, a series of new revolutionary technologies defined by higher frequencies and extremely large-scale multiple-input multiple-output (MIMO) are proposed. Specifically, MIMO, due to its ability in rendering spatial multiplexing and beamforming gains, plays a pivotal role in both 5G and 6G wireless networks. Nevertheless, traditional MIMO approaches focused mainly on adapting to existing wireless conditions through transceiver design, relying on \emph{passive alignment} with the wireless environment. Recently, cutting-edge technologies have begun to disrupt this paradigm, offering new possibilities to \emph{actively reshape and customize} the wireless propagation environment, marking a significant shift from passive adaptation to proactive intervention in signal transmission. In particular, two promising yet different technologies emerging from the field of antenna design---Reconfigurable Intelligent Surfaces (RIS)\cite{QQProc2024,wang2021Joint,ning2021terahertz} and Movable Antenna (MA)/Fluid Antenna (FA)\cite{ZhuMa25,NingMov25}---have garnered growing interest, owing to their unique capabilities in dynamically reshaping the signal propagation environment. In particular, RIS, comprising a large number of nearly passive reflecting elements, manipulates the amplitude and phase of impinging signals, thereby allowing the reflected waves to combine constructively or destructively at the intended receiver \cite{QQProc2024,wang2021Joint}. On the other hand, MA and FA reconfigure wireless channels by dynamically adjusting the physical attributes of antennas, such as their location, orientation, or shape, through mechanical or electronic devices\cite{ZhuMa25,NingMov25,KKWong21,shao20256dma}.

Although RIS and MA differ in terms of antenna design and hardware implementation, they both reconfigure the wireless channel by adjusting real-time controllable coefficients or parameters. To illustrate this operation, let $\boldsymbol{H}(\mathcal{A};\mathcal{C})$ denote the end-to-end wireless channel, in which $\mathcal{A}$ represents adjustable coefficients and $\mathcal{C}$ denotes channel propagation parameters that are determined by environmental scatterers and geometries. Thus, the end-to-end wireless channel is a function of both the adjustable coefficients $\mathcal{A}$ and the propagation parameters $\mathcal{C}$. Fig. \ref{Fig-H1} (a) and (b) illustrate the resulting end-to-end channels for RIS-assisted and MA-enabled single-input single-output (SISO) systems, respectively.


Ideally, we would like to optimize the controllable coefficients $\mathcal{A}$ to obtain a favorable end-to-end wireless channel $\boldsymbol{H}$. More specifically, for MIMO systems, we are particularly interested in jointly designing the adjustable coefficients $\mathcal{A}$ and the transceiver beamforming variable $\mathcal{F}$ to achieve a maximum spectral efficiency (SE) or optimize other performance metrics. Let $f(\boldsymbol{H}(\mathcal{A};\mathcal{C}), \mathcal{F})$ denote the objective function such as the SE or the received signal power, which often admits an explicit expression in terms of $\boldsymbol{H}(\mathcal{A};\mathcal{C})$ and $\mathcal{F}$. The optimization problem can then be formulated as
\begin{align}
	\max_{\mathcal{A},\mathcal{F}} \; f(\boldsymbol{H}(\mathcal{A};\mathcal{C}), \mathcal{F}). \label{eq-f}
\end{align}



To address this problem, the majority of existing studies presuppose an explicit expression of $\boldsymbol{H}(\mathcal{A};\mathcal{C})$ and proceed by estimating the underlying propagation parameters $\mathcal{C}$. In doing so, the problem in \eqref{eq-f} reduces to a white-box optimization problem, wherein conventional optimization methods can be employed.



However, exact modeling of $\boldsymbol{H}(\cdot)$ maybe challenging as it requires accurate electromagnetic characterisation of the reconfigurable materials/devices as well as accurate modeling of the complex interaction between the reconfigurable devices and the electromagnetic waves. Additionally, the accuracy of the channel model of $\boldsymbol{H}(\cdot)$ is also affected by hardware imperfections inherent in the reconfigurable devices such as mutual coupling between adjacent antennas/elements. In addition, obtaining the environmental propagation parameters $\mathcal{C}$ is often a formidable task. For RIS-assisted systems, obtaining $\mathcal{C}$ necessitates addressing the nearly passive nature of the RIS, and usually incurs a prohibitively high training overhead due to the massive number of reflecting elements. For MA-enabled systems, acquiring $\mathcal{C}$ means obtaining the angular/delay/attenuation parameters associated with each signal path from the transmitter to the receiver. This process demands substantial training overhead in rich-scattering environments and is vulnerable to model mismatches.

The crux of traditional methods lies in their dependence on knowledge of the explicit expression $\boldsymbol{H}(\mathcal{A},\mathcal{C})$ and the propagation parameters $\mathcal{C}$, which is essential for computing gradient information w.r.t. $\mathcal{A}$ and $\mathcal{F}$ in the subsequent joint beamforming design. To address this issue, we consider a plug-and-play framework for channel reconfiguration without relying on the knowledge of $\boldsymbol{H}(\cdot)$ and $\mathcal{C}$. We do so by introducing derivative-free optimization, which treats end-to-end channel reconfiguration/optimization as a black-box problem. The motivation behind this approach stems from in a key observation: evaluating the \emph{value} of the objective function $f(\cdot)$ is often more tractable than modelling $\boldsymbol{H}(\cdot)$ or estimating $\mathcal{C}$ from the received signals. Consider the RIS-assisted SISO system illustrated in Fig. \ref{Fig-H1}(a) as an example, and let $f(\cdot)$ represent the received signal power. By having the transmitter send pilot signals while keeping the reflection coefficients (i.e., $\mathcal{A}$) fixed, the received signal power can be readily computed from the received signals, without requiring knowledge of the two-hop channels or the propagation parameters $\mathcal{C}$. By evaluating the objective function values derived from received signals, the approach employs function queries to \emph{approximate} gradient information w.r.t. $\mathcal{A}$ and $\mathcal{F}$. This estimated gradient information facilitates the advanced optimization techniques to solve \eqref{eq-f}. Owing to its computational simplicity and competitive practical performance, this approach has demonstrated significant efficacy in many applications such as black-box adversarial attacks and language models fine-tuning\cite{LiuZO20,malladi2023fine}. Nevertheless, its potential for signal processing in reconfiguring wireless communications remains largely unexplored.

This article makes a first attempt to investigate the potential of derivative-free optimization in channel reconfiguration enabled by RIS and MA/FA technologies. We begin by presenting the fundamentals of derivative-free optimization and outlining a general framework for its application to channel reconfiguration problems. In this context, derivative-free optimization is considered a powerful tool for directly reshaping the propagation environment, without necessitating either explicit knowledge of $\boldsymbol{H}(\cdot)$ or $\mathcal{C}$. Then, we provide two case studies to demonstrate the superiority of the derivative-free optimization method over state-of-the-art approaches in addressing the passive beamforming problem for RIS-assisted systems and the antenna position optimization problem for MA/FA systems, respectively. Finally, we discuss key opportunities and future research challenges, followed by concluding remarks.

\section{Principles and Applications of Derivative-Free Optimization}
This section begins by introducing the principles of derivative-free optimization and then explores its application to the challenging problem of channel reconfiguration.

\subsection{Principles of Derivative-Free Optimization}
Derivative-free optimization, which refers to methods that do not require the availability of derivatives, has attracted significant research attention from the optimization, signal processing, and machine learning communities over the past decade \cite{LiuZO20,malladi2023fine}. In particular, zeroth-order (ZO) methods stand out among other approaches due to their elegant theoretical guarantees and fast convergence\cite{LiuZO20}. The most prominent feature
of ZO methods is that they only rely on \emph{values} of the
objective function in lieu of the exact gradient w.r.t. optimization variables. This characteristic renders ZO methods indispensible in scenarios where the objective function depends on unknown or intractable parameters, yet its values can be directly accessed through observations. As previously stated, the values of the objective function $f(\cdot)$ at the receiver is readily obtainable in channel reconfiguration problems. Specifically,
by letting the transmitter send pilot symbols, the receiver can obtain an estimate of the end-to-end channel $\boldsymbol{H}(\mathcal{A}^{(t)};\mathcal{C})$ for a given $\mathcal{A}^{(t)}$. It is worth noting that the function $f(\cdot)$ can be explicitly evaluated once the end-to-end channel $\boldsymbol{H}(\mathcal{A};\mathcal{C})$ and the transceiver beamforming $\mathcal{F}$ are specified. Thus, the receiver can easily calculate the objective function value $f(\boldsymbol{H}(\mathcal{A}^{(t)};\mathcal{C}),\mathcal{F}^{(t)})$ attained by the current set of adjustable coefficients $\mathcal{A}^{(t)}$ and the currently used transceiver beamforming variable $\mathcal{F}^{(t)}$.

Such a convenience of accessing the objective function value at the receiver makes ZO methods extremely
eligible for end-to-end channel reconfiguration without the explicit knowledge of $\mathcal{C}$ and expression of $\boldsymbol{H}(\mathcal{A};\mathcal{C})$.
To elucidate ZO methods for our problems, we first
provide an introduction to the basic principles of zeroth-order
optimization.

\begin{figure}
\centering
\includegraphics[width=3.6in]{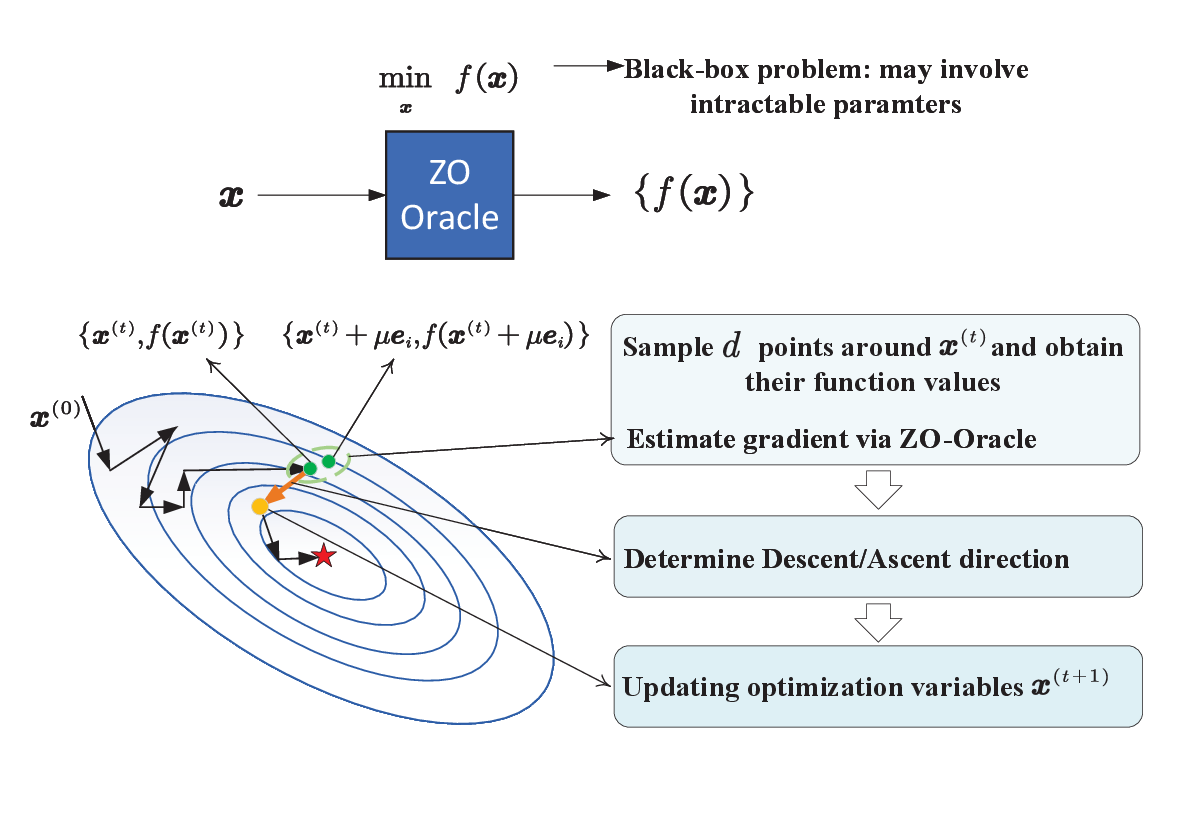}
\caption{Illustration of the iteration process of ZO methods to solve a black-box optimization problem: The $t$th iteration usually involves three steps. Firstly, the objective functional values of points at a set of points $\{ \boldsymbol{x}^{(t)} + \mu \boldsymbol{e}_i\}_{i=1}^{d}$ in the vicinity of $\boldsymbol{x}^{(t)}$ are evaluated and used to approximate the gradient at $\boldsymbol{x}^{(t)}$ according to \eqref{ZO-gradient}. Second, based on the estimated gradient, a descent/ascent direction is determined using techniques in first-order optimization methods. Finally, the next sampling point $\boldsymbol{x}^{(t+1)}$ is obtained by updating $\boldsymbol{x}^{(t)}$ along the chosen direction. This process is repeated until convergence to a critical point. }
	\label{fig-ZO-update}
\end{figure}

The pivotal idea of ZO is to employ a ZO gradient to replace
the exact gradient that is required by first-order optimization methods.
Typically, a widely adopted approach of calculating the ZO
gradient is given as
\begin{align}
\hat{\nabla}f(\boldsymbol{x})\triangleq \frac{1}{\mu}
\sum_{i=1}^d \Big( f(\boldsymbol{x}+\mu\boldsymbol{e}_i)
-f(\boldsymbol{x})\Big)\boldsymbol{e}_i,
\label{ZO-gradient}
\end{align}
where $f(\boldsymbol{x})$ is the objective function,
$\boldsymbol{x}$ is the $d$-dimensional variable vector corresponding to $\mathcal{A}$ and $\mathcal{F}$,
$\boldsymbol{e}_i$ is the $i$th canonical basis vector,
and $\mu$ is a parameter controlling the error
between the ZO gradient and the exact gradient.
Essentially, the form of $\hat{\nabla}f(\boldsymbol{x})$
is to let each coordinate of $\hat{\nabla}f(\boldsymbol{x})$
be chosen as the corresponding directional derivative.
It is easy to see that $\hat{\nabla}f(\boldsymbol{x})$
becomes the exact gradient when $\mu$ tends to $0$. In the
light of (\ref{ZO-gradient}), computing the ZO gradient
$\hat{\nabla}f(\boldsymbol{x})$ only requires the values
of $f(\boldsymbol{x}+\mu\boldsymbol{e}_i)$ and
$f(\boldsymbol{x})$. When the ZO gradient is obtained, a
gradient ascent step can be performed to update the
variable $\boldsymbol{x}$, as shown in Fig. \ref{fig-ZO-update}.

Albeit simple and straightforward, ZO methods offer significant advantages. Firstly, the inherent function evaluation-based mechanism bypasses the need to capture the global picture of $\boldsymbol{H}(\cdot)$. Next, the sample complexity of ZO methods scales linearly with the dimensionality of the optimization problem\cite{RiemICML}, making them more efficient in high-dimensional scenarios compared to reinforcement learning (RL)-based methods, which incur exponential sample complexity \cite{sutton2018reinforcement}. Moreover, ZO methods involve only a single step-size hyperparameter, which is well-studied in first-order optimization. These characteristics make ZO methods practical and efficient for dynamic and reconfigurable wireless communication systems, offering superior efficiency, scalability, and robustness compared to RL.

\begin{figure*}[!thbp]
\centering
\includegraphics[width=7in]{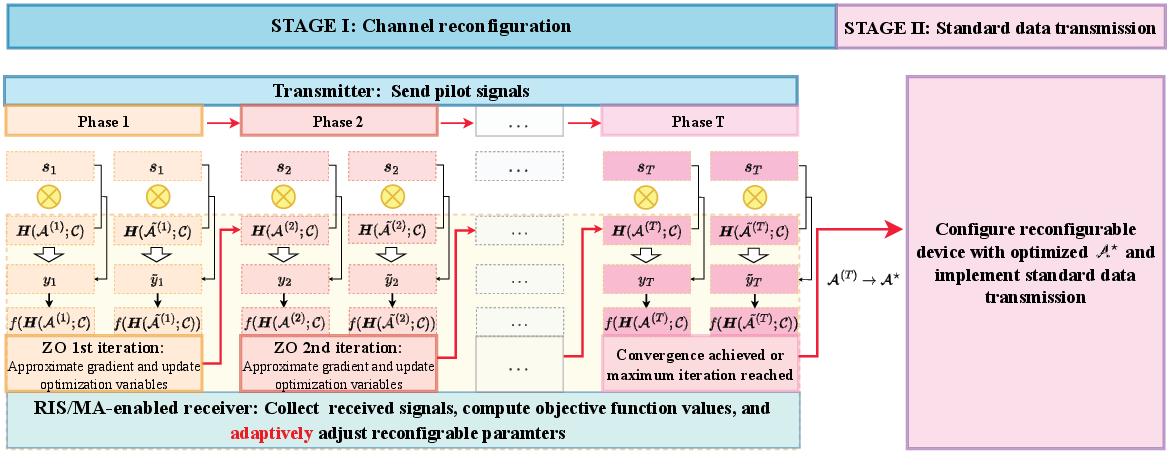}
\caption{A general framework for applying ZO methods in RIS/MA-enabled reconfigurable environments: The framework comprises a channel reconfiguration stage and a standard data transmission stage. Stage I consists of $T$ phases, each corresponding with one iteration of the ZO method. In the $t$th iteration, the transmitter sends a pilot signal $s_t$. The reconfigurable device is sequentially configure with $\mathcal{A}^{(t)}$ and a slightly perturbed version $\mathcal{\tilde{A}}^{(t)} = \mathcal{A}^{(t)} + \mu \boldsymbol{e}_i$. The pilot signals are then propagated through the end-to-end channels $\boldsymbol{H}(\mathcal{A}^{(t)};\mathcal{C})$ and $\boldsymbol{H}(\tilde{\mathcal{A}}^{(t)};\mathcal{C})$, respectively. Based on received signals $y_t$ and $\tilde{y}_t$, the receiver can compute the objective function values $f(\boldsymbol{H}(\mathcal{A}^{(t)};\mathcal{C}))$ and $f(\boldsymbol{H}(\tilde{\mathcal{A}}^{(t)};\mathcal{C}))$, respectively. Through function evaluations, the ZO method approximates the gradient and update the optimization variables to obtain $\mathcal{A}^{(t+1)}$. The process continues until a stopping criterion is met. The final output $\mathcal{A}^{(T)}$ is taken as $\mathcal{A}^{\star}$ and fixed for subsequent standard data transmission stage.  }
\label{fig-scheme}
\end{figure*}




\subsection{Applications in Channel Reconfigurable Problems}
Herein, we discuss how to apply ZO methods to address the channel reconfiguration problem without assuming the knowledge of $\mathcal{C}$ and the expression of $\boldsymbol{H}(\mathcal{A}; \mathcal{C})$. To facilitate our exposition, we ignore the transceiver beamformer $\mathcal{F}$ and simply consider the problem of optimizing the reconfigurable coefficients: $\max_{\mathcal{A}}  f(\boldsymbol{H}(\mathcal{A}, \mathcal{C}))$. Note that the principle of ZO can be readily adapted to the more general joint transceiver design and channel reconfiguration problem.

To optimize the variable $\mathcal{A}$, ZO methods sequentially samples and updates $\mathcal{A}$ by leveraging the ZO gradient derived solely from the objective function values. Specifically, we can let the receiver coordinate with the reconfigurable device (e.g., the RIS) to efficiently implement the ZO method. Fig. \ref{fig-scheme} illustrates a general framework of the derivative-free optimization-based channel reconfiguration process. In specific, the proposed framework consists of a channel reconfiguration stage and a standard data transmission stage within a coherence time interval where the environment parameters $\mathcal{C}$ are assumed to remain invariant. In stage I, the transmitter sends pilot signals. For the current set of reconfigurable coefficients $\mathcal{A}^{(t)}$, the receiver, based on the received pilot signals, can evaluate the corresponding objective function value $f(\boldsymbol{H}(\mathcal{A}^{(t)}; \mathcal{C}))$. Note that, to calculate the ZO gradient \eqref{ZO-gradient}, the receiver has to calculate the objective function values at multiple (at least two) sampling points of $\mathcal{A}$. Such a process can be easily implemented by letting the receiver coordinating with the reconfigurable device. After the ZO gradient at the current point $\mathcal{A}^{(t)}$ is obtained, a gradient ascent step can be performed to determine the next sampling point $\mathcal{A}^{(t+1)}$. This iterative procedure continues until either the maximum number of iterations is reached or the convergence is achieved. In stage II, with the optimized reconfigurable coefficients, standard data transmission is then implemented.

The above framework effectively unifies the two conventional phases, i.e., acquisition of propagation parameters $\mathcal{C}$ (e.g., channel estimation) and reconfigurable coefficients optimization, into a single channel reconfiguration stage. By adaptively refining the optimization variable $\mathcal{A}$, the RIS/MA-enabled receiver can realize a favorable wireless propagation environment in a self-adaptive manner, relying solely on the received signals. This obviates the need for explicit modeling of $\boldsymbol{H}(\mathcal{A}, \mathcal{C})$ and the acquisition of the propagation parameters $\mathcal{C}$. In the following section, we will delve into two specific case studies to demonstrate the potential of ZO methods in dynamic wireless channel reconfiguration.

\section{Case Studies}
\begin{figure*}[!htbp]
\centering
\includegraphics[width=7.2in]{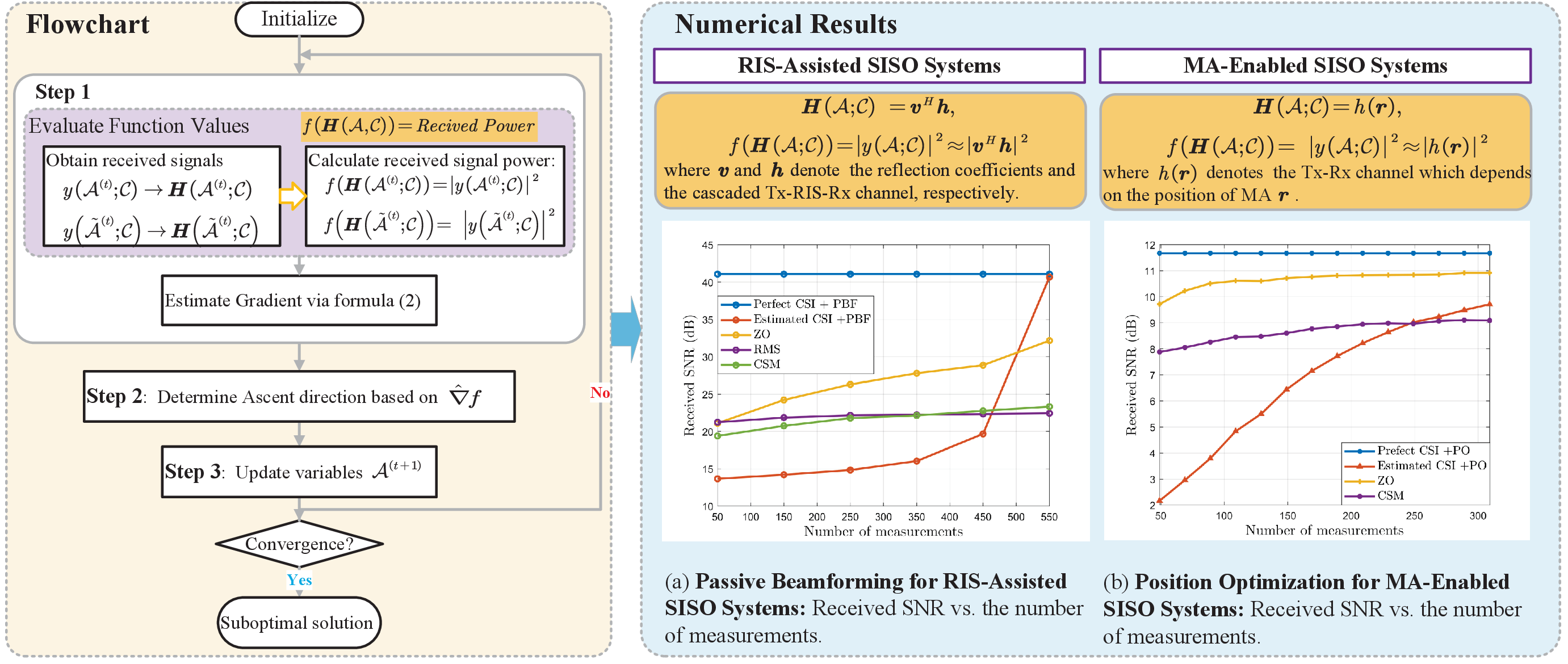}
\caption{Algorithm flowchart and numerical performance of applying ZO methods in channel reconfiguration in RIS- and MA-enabled SISO systems: The left panel illustrates the iterative procedure of the ZO method, which consists of three steps: (1) objective function evaluation and gradient approximation, (2) ascent direction determination, and (3) update of optimization variables. These three steps correspond directly to the steps depicted in Fig.2. The objective function $f(\cdot)$ is set to as the received signal power, which can be readily computed from received signals. The right panel presents numerical results demonstrating the effectiveness of the ZO method in (a) passive beamforming for RIS-assisted systems and (b) antenna position optimization for MA-enabled systems.}
\label{fig-ZOresults}
\end{figure*}

To illustrate the effectiveness of ZO methods, this section presents two case studies that demonstrate both the conceptual algorithm flowchart and numerical performance of applying ZO methods for channel reconfiguration, as depicted in Fig. \ref{fig-ZOresults}. The first case study focuses on optimizing the reflection coefficients of RIS, while the second aims to find the optimal position for a MA.
We compare the ZO method with the classical optimization approach which first estimates the environmental propagation parameters $\mathcal{C}$ and subsequently optimizes the controllable coefficients to maximize the objective function.

\subsection{Passive Beamforming for RIS-Assisted SISO Systems}
We first consider a RIS-assisted single-input single-output (SISO) system, where the RIS is equipped with $M=512$ reflection elements. The objective is to maximize the received signal power while satisfying the unit modulus constraint imposed on the RIS reflection coefficients. In this example, the propagation parameter $\mathcal{C}$ refers to the cascaded transmitter-RIS-receiver channel.

The performance of the ZO method is compared against existing state-of-the-art methods. The first, termed ``Perfect CSI + passive beamforming (PBF)", optimizes the reflection coefficients to maximize the received signal power based on the perfect knowledge of $\mathcal{C}$, i.e., the channel state information (CSI) of the cascaded channel. Clearly, this approach serves as a performance upper bound for all methods. The second method, referred to as ``Estimated CSI + PBF" utilizes a least-squares method to estimate the cascaded channel and subsequently optimize the reflection coefficients based on the estimated CSI. In addition, other beamforming techniques without requiring the knowledge of $\mathcal{C}$, namely, the random-max sampling (RMS) method and the conditional sample mean (CSM) method, are also included for comparison. The RMS method randomly generates a set of reflection coefficient vectors and configures RIS with these vectors to obtain the corresponding received measurements. Then, RMS selects the one that maximizes the received signal power. The CSM method determines the reflection coefficient $\theta_n$ that maximizes the total received signal power conditioned on $\theta_n$ \cite{ren2023Configuring}. The CSM method is known to achieve the well-established quadratic scaling law as the number of measurements approaches infinity. Notably, the CSM method can also be interpreted as solving a \emph{multi-armed bandit problem in RL} \cite{lai2024Blind}. In this experiment, the resolution of the RIS phase shifter is set to $2$ to match the design constraints of the CSM method.


Fig. \ref{fig-ZOresults}(a) illustrates the received signal-to-noise ratios (SNRs) achieved by respective methods as a function of the number of pilot symbols. It is seen that the ``Estimated CSI + PBF'' method performs poorly when the number of pilots is fewer than the number of reflecting elements, in which case the cascaded channel cannot be properly estimated due to the insufficient number of measurements. This limitation significantly impacts its scalability to larger RIS-enabled MIMO systems. Conversely, the ZO method optimizes reflection coefficients directly without requiring the knowledge of the cascade channel. We observe that the ZO method presents a substantial performance improvement over the ``Estimated CSI + PBF'' method in the regime of insufficient samples. This result demonstrates the sample efficiency of the ZO method in solving the channel reconfiguration optimization problem. It also outperforms the other two beamforming methods, i.e., RMS and CSM, by a big margin when a moderate number of pilot symbols are available.
The superior performance of the ZO method over the CSM method can be attributed to its intelligent sampling strategy. While the CSM method relies on finding the statistical relationship between the objective function value and each reflection coefficient, the ZO method dynamically determines its sampling trajectory based on gradient approximations. This adaptive sampling mechanism allows the ZO method to converge more quickly and efficiently toward the optimal reflection coefficients, yielding a superior tradeoff between the performance and the sample complexity.


\subsection{Position Optimization for MA-Enabled Systems}
Next, we consider a point-to-point scenario where a single MA is deployed at the receiver, while the transmitter is equipped with a fixed-position antenna. The objective is to optimize the position of the receive MA to maximize the received signal power, subject to a constraint placed on the movable region \cite{ZengFang25}. The movable region for the MA is defined as a square area of size $2\lambda \times 2\lambda$, where $\lambda$ denotes the carrier wavelength.
In this example, the propagation parameters $\mathcal{C}$ refer to the CSI including the angle, the attenuation, and the delay parameters of each signal path between the transmitter and the receiver. Such information is sufficient to reconstruct the wireless channel for any position of the receiver antenna. Similarly, traditional optimization methods and the RMS method are considered for comparison. Traditional approaches, including the ``Perfect CSI + Position Optimization (PO)'' and ``Estimated CSI + PO'' optimize the position of the receive antenna based on perfect or estimated CSI. The former provides a performance upper bound while the latter utilizes compressive sensing (CS)-based approaches to estimate the propagation parameters $\mathcal{C}$ by exploiting its sparse structure.



Fig. \ref{fig-ZOresults}(b) illustrates the received SNRs of respective methods as a function of the number of pilot symbols. The ``Estimated CSI + PO'' method exhibits a sharp increase in the SNR initially, followed by performance saturation as the number of measurements increases. The ZO-based method consistently outperforms both the RMS method and the ``Estimated CSI + PO'' method, while experiencing only a slight performance loss as compared to the upper bound achieved by ``Perfect CSI + PO''. This highlights the superior sample efficiency of the ZO method. Such a sample efficiency enables the
ZO algorithm to use very limited samples to rapidly reconfigure the end-to-end channel, which is highly desirable in fast-changing channel
environments. It should be noted that the performance of ``Estimated CSI + PO'' method relies on exploiting the sparse structure of the field response channel model. However, in rich-scattering environments, the number of resolvable paths may be large. This renders the sparse model assumption invalid, and thus the CS-based approach less effective or even infeasible. In contrast, the ZO method does not rely on specific channel models. Its robustness to model mismatches makes it a practical and efficient solution for MA-enabled systems.

In summary, the results from both case studies demonstrate the clear superiority of the ZO method over existing approaches. Table \ref{table-1} presents several key advantages of ZO-based methods compared to existing methods. These findings confirm the great potential of the ZO method to effectively address optimization challenges in reconfigurable wireless environments.

\begin{table*}[!ht]
\centering
\caption{Key Advantages of ZO-based Methods over Existing Methods.}
\begin{tabular}{c}
 \includegraphics[width=0.8\textwidth]{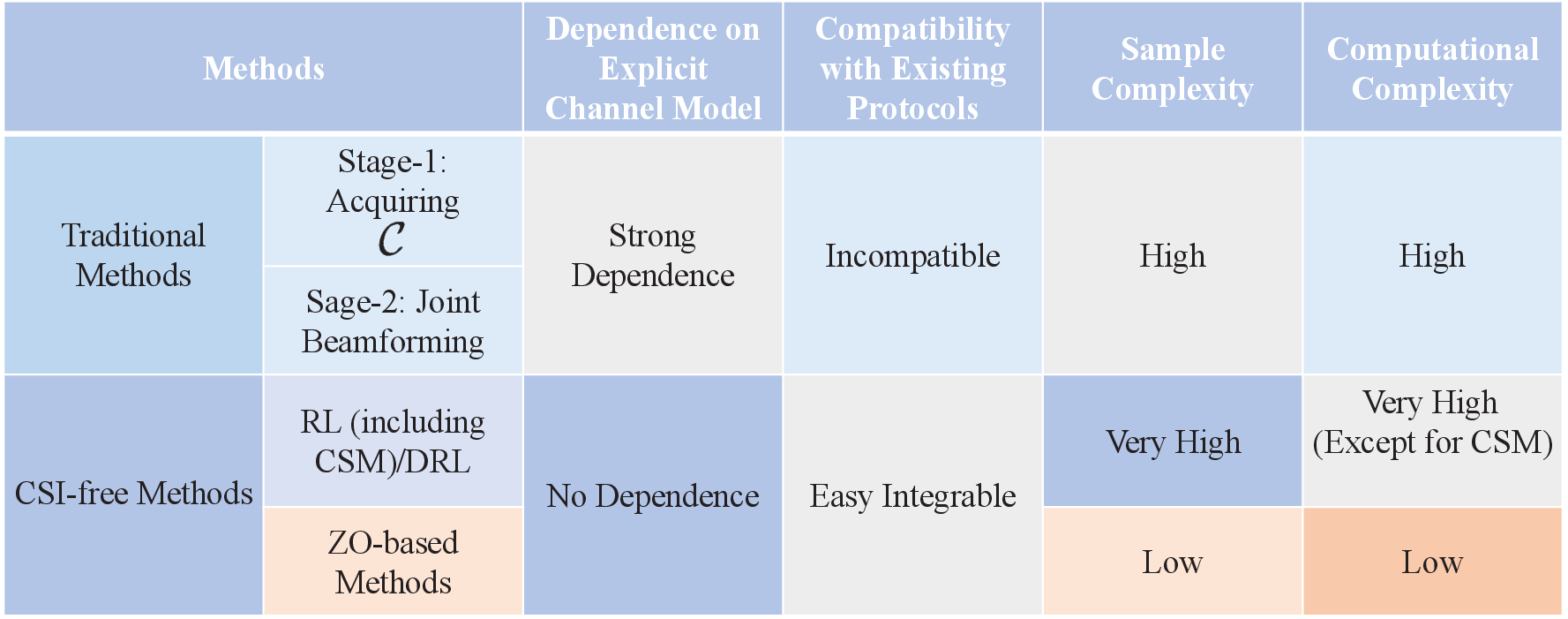} \\
\end{tabular}
\label{table-1}
\end{table*}

\section{Opportunities of ZO in Reconfigurable Systems}
Thanks to the simple and elegant form of ZO optimization (ZOO), applying it to reconfigurable systems offers considerable advantages. In this section, we outline several key benefits.
\subsection{Powerful tool to combat unknown environments}
As indicated in Section II, ZOO does not require any knowledge of the channel model $\boldsymbol{H}(\cdot)$ or the propagation parameters $\mathcal{C}$. Instead, all it needs is the objective function value, e.g., the received signal power, to directly optimize the reconfigurable coefficients. This feature makes it a powerful tool for addressing the challenges of acquiring $\boldsymbol{H}(\cdot)$ and $\mathcal{C}$ in both RIS-assisted and MA-enabled systems. As explained in Section I, obtaining these parameters demands significant time and computational resources. In cases where acquiring $\boldsymbol{H}(\cdot)$ and $\mathcal{C}$ is infeasible, such as in physical layer security scenarios where non-cooperative jammers or eavesdroppers attempt to attack legitimate transmissions, ZOO may be the only effective method to enhance the overall system performance.
\subsection{Improved Sample Complexity}
ZOO addresses the black-box optimization problem by adaptively sampling the feasible space using a gradient-approximation strategy, akin to gradient-descent methods. This gradient-based sampling strategy enables ZOO to efficiently exploit the local characteristics of the objective function around the sampling point. Consequently, the sample complexity of ZOO is significantly improved compared to traditional derivative-free optimization methods, such as Bayesian optimization. As indicated in Section II.B, the convergence analysis of ZOO reveals that its sample complexity grows linearly with the dimensionality of the optimization problem\cite{RiemICML}, whereas the DRL approach and other derivative-free optimization methods may grow exponentially. This makes ZOO a much more practical solution for optimization in reconfigurable systems with large-scale reflecting elements or movable antennas.
\subsection{Compatibility with existing protocols}
ZOO enables RIS and MA-enabled receivers to operate in a plug-and-play fashion within reconfigurable propagation environments. First, RIS and MA-enabled receivers can function independently, meaning the service provider (Tx) does not need to coordinate with them or even be aware of their presence. Specifically, RIS and MA-enabled receivers can adaptively adjust their reconfigurable coefficients based on previously received pilots or data. Furthermore, the adjustment process can be terminated once the improvement in the objective function value satisfies a predefined constraint. For example, in RIS-assisted systems aiming to maximize the received signal power, the adjustment of reflection coefficients can be stopped once the received signal power exceeds a certain threshold. This feature makes ZOO highly compatible with existing protocols.

\section{Future Directions}
In this section, we list several key future directions for advancing the derivative-free optimization techniques for reconfigurable wireless communication systems.

\subsection{ZOO with Various Hardware Constraints}
Due to hardware constraints, in reconfigurable systems, the controllable coefficients cannot take arbitrary values and are often subject to a variety of nonconvex constraints such as unit modulus constraints and discrete-value constraints. For example, the magnitude of the reflection coefficients for RIS is typically subject to a unit modulus constraint. Similarly, in pixel-enabled MA or FA systems, the antenna position has to be discretized due to hardware limitations. For such hardware constraints, the reconfigurable coefficients/variables must always satisfy these restrictions throughout the sequential query process. The discontinuity of the feasible space can lead to a slow convergence rate or even divergence of ZO methods. Thus, further efforts are expected to deepen the convergence analysis and extend ZO methods to more general and complex scenarios.

\subsection{ZOO with Imperfect Query Measurements}
Typical ZO methods in machine learning rely on the assumption of perfect query measurements. However, in reconfigurable wireless systems, measurements are often obtained as functions of the received signal, which are inevitably corrupted by noise and errors due to distortion in wireless propagations. These imperfect measurements may significantly impact the convergence of ZO methods, leading to performance degradation in the system. Robustness analysis of ZO methods under such noise-corrupted measurements remains an open challenge. Addressing this gap is particularly important for wireless communication scenarios, where signal noise is pervasive and poses unique difficulties compared to traditional machine learning contexts.

\subsection{ZOO with Accelerated Tricks}
Although the ZOO has demonstrated significant potential in reducing the sample complexity compared to the other closed-box optimization methods, further acceleration of ZOO remains appealing especially for extremely large-scale MIMO systems. Typically, there is a trade-off between \emph{the number of query points per iteration}, i.e, $d$, and \emph{the number of iterations} to reach a given precision. Increasing $d$ may improve the gradient estimation accuracy and require less iterations to converge, while fewer query points per iteration may increase the number of iterations needed for convergence.
Identifying methods that enhance the accuracy and robustness of gradient estimation is crucial for further improving the sample efficiency of ZOO. Such advancements could significantly improve its applicability to reconfigurable antenna-empowered systems.

\section{Conclusions}
This paper discussed significant challenges faced by existing methods when addressing channel reconfiguration problems and introduced ZOO as a powerful tool to overcome these challenges. The fundamental principles of ZOO were presented, highlighting its potential advantages over RL approaches, particularly emphasizing the suitability of its simple function value evaluation property for solving channel reconfiguration problems. Two illustrative case studies involving RIS and MA clearly demonstrated the superiority of ZOO-based methods. Finally, several promising future research directions were outlined to further enhance the application of ZOO methods in reconfiguring wireless propagation environments.

\bibliography{newbib}
\bibliographystyle{IEEEtran}

\end{document}